\documentclass[aps,prb,12pt,citeautoscript,superscriptaddress]{revtex4-1}
\usepackage{amsmath,amssymb,mathrsfs,bm,setspace}
\usepackage{graphicx}
\usepackage{color}
\usepackage[colorlinks=true]{hyperref}
\hypersetup{
    bookmarks=true,         % show bookmarks bar?
    unicode=false,          % non-Latin characters
    pdftoolbar=true,        % show Acrobat
    pdfmenubar=true,        % show Acrobat
    pdffitwindow=false,     % window fit to page when opened
    pdfstartview={FitH},    % fits the width of the page to the window
    pdftitle={My title},    % title
    pdfauthor={Author},     % author
    pdfsubject={Subject},   % subject of the document
    pdfcreator={Creator},   % creator of the document
    pdfproducer={Producer}, % producer of the document
    pdfkeywords={keyword1} {key2} {key3}, % list of keywords
    pdfnewwindow=true,      % links in new window
    colorlinks=true,       % false: boxed links; true: colored links
    linkcolor=red,          % color of internal links (change box color with linkbordercolor)
    citecolor=blue,        % color of links to bibliography
    filecolor=magenta,      % color of file links
    urlcolor=cyan           % color of external links
}
\usepackage{bm}
\usepackage{latexsym}
\usepackage[section]{placeins}

\def \1{\bf 1}
\def \2{\bf 2}

\newcommand{\dd}{\mathrm{d}}
\newcommand{\ii}{\mathrm{i}}
\newcommand{\beq}{\begin{equation}}
\newcommand{\eeq}{\end{equation}}
\newcommand{\ba}{\begin{array}{ccc}}
\newcommand{\ea}{\end{array}}
\newcommand{\nn}{\nonumber \\}
\def\bea{\begin{eqnarray}}
\def\eea{\end{eqnarray}}

\begin{document}

\title{Spectral function of a localized fermion\\ coupled to the Wilson-Fisher conformal field theory}
 \author{Andrea Allais}
 \affiliation{Department of Physics, Harvard University, Cambridge, MA 02138, USA}
 \author{Subir Sachdev}
 \affiliation{Department of Physics, Harvard University, Cambridge, MA 02138, USA}
\affiliation{Perimeter Institute for Theoretical Physics, Waterloo, Ontario N2L 2Y5, Canada}
 \date{\today\\
}
 %\vspace{1.6in}
\begin{abstract}
We describe the dynamics of a single fermion in a dispersionless band coupled to the 2+1 dimensional 
conformal field theory (CFT) describing the quantum phase transition of a bosonic order parameter with $N$ components.
The fermionic spectral functions are expected to apply to the vicinity of quantum critical points in two-dimensional metals 
over an intermediate temperature regime where the Landau damping of the order parameter can be neglected.
Some of our results are obtained by a mapping to an auxiliary problem of a CFT containing a defect line with an external field 
which locally breaks the global O($N$)
symmetry.
\end{abstract}
\maketitle
%\tableofcontents

\section{Introduction}
\label{sec:intro}

It is well-known that the Wilson-Fisher conformal field theory (CFT) describes the quantum phase transition
of a number of boson and insulating spin models.\cite{chn,fwgf,csy,coldea,sstilt}
In the presence of the Fermi surface of metals, the order parameter quantum fluctuations undergo Landau-damping, and there is a crossover to a low energy regime controlled by the physics of the Fermi 
surface \cite{hertz,millis,vrmp,AC04,met1,met2,mross,strack,dalidovich,sur}.
However, it is possible that the magnitude of the Landau damping is parametrically small,\cite{sokol,georges,kachru1,kachru2,raghu}
and then there is a significant intermediate energy regime over which the fermions are
coupled to the `relativistic' ({\em i.e.\/} with dynamic critical exponent $z=1$) order parameter dynamics of the
Wilson-Fisher CFT. It is this intermediate energy regime\cite{sokol,georges} which is the focus of attention of 
the present paper. 

A key feature of the dynamics of fermions coupled to Wilson-Fisher bosons is the renormalization group flow
of the fermion dispersion. When the fermions have a quadratic dispersion, it is clear that there is a flow to 
a flat, dispersionless fermion band.\cite{troyerprl} For the case of fermions with a non-zero Fermi velocity, $v_F$, 
Fitzpatrick {\em et al.\/} \cite{kachru2} have recently argued that the flow to a flat band with $v_F \rightarrow 0$ persists. 
So for a discussion of the intermediate energy regime noted above, we are therefore led to consider 
the problem of a dispersionless band of fermions interacting with bosonic degrees of freedom
in two spatial dimensions described by the Wilson-Fisher fixed point.

\begin{figure}
\includegraphics[scale=0.5]{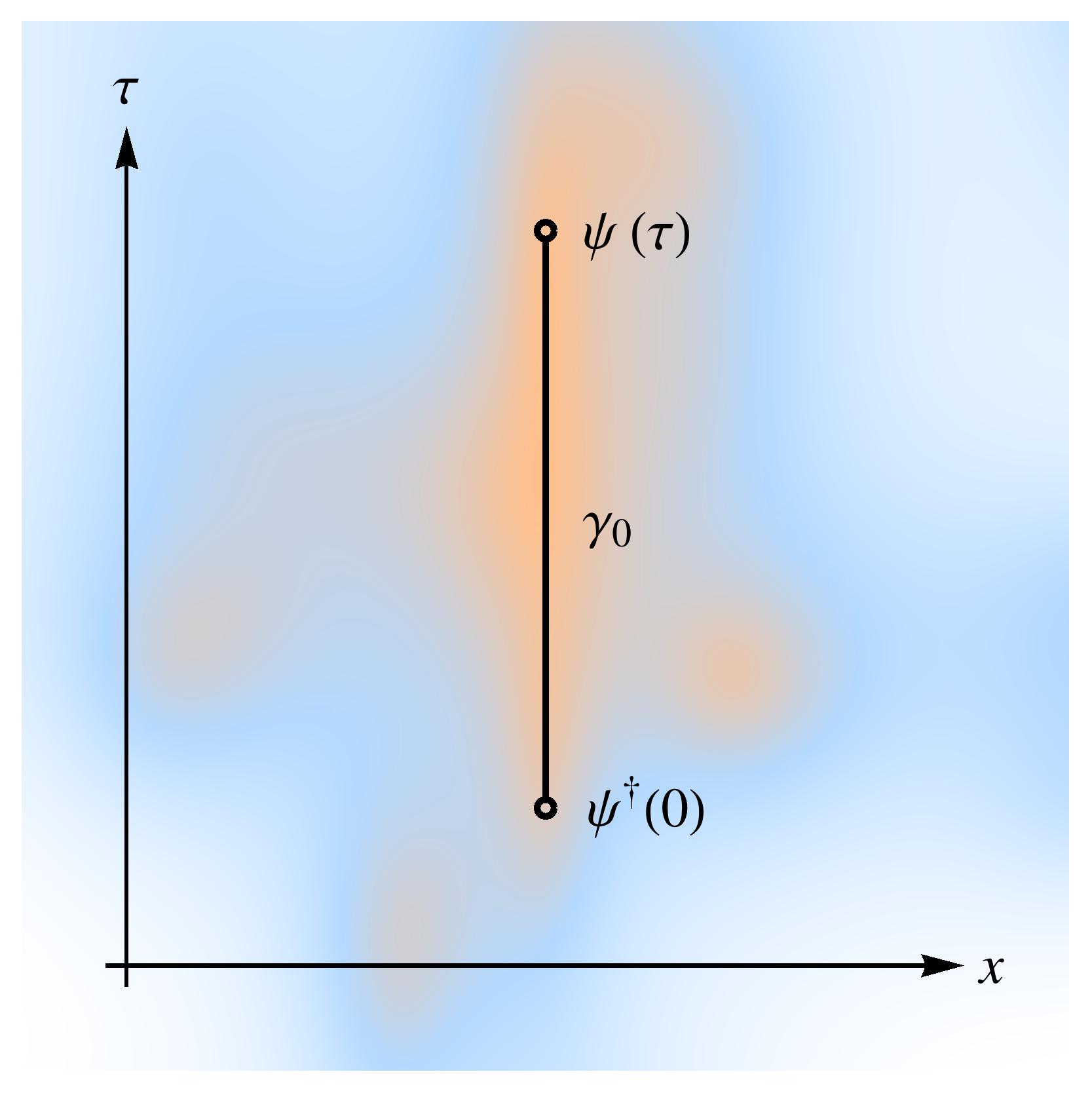}  
\caption{Spacetime representation of the theory $\mathcal{Z}$. The Wilson-Fisher CFT degrees of freedom extend over all spacetime.
The fermion is created at time 0 and annihilated at time $\tau$. While the fermion is present, a local field $\gamma_0$ acts on the
CFT degrees of freedom at $x=0$ along the direction $\alpha = 1$.
}
\label{fig:spacetime}
\end{figure}

We can now make further simplifications for the field-theoretic critical 
analysis of this limiting fermion-boson problem. As we will be ignoring
the Landau damping arising from particle-hole loop diagrams of fermions, 
we may as well take only a single fermion in the dispersionless band.\cite{troyerprl,ssimpurity} Furthermore, because this fermion is 
dispersionless, we are free to localize it \cite{ssimpurity} at a single spatial point $x=0$.
We are therefore led to consider the following partition function of a single fermion $\psi (\tau)$ coupled to the 
$N$-component order parameter $\phi_{\alpha} (x, \tau)$ ($\alpha = 1 \ldots N$) of the Wilson-Fisher theory in $d$ spatial
dimensions ($x$) and one imaginary time ($\tau$) dimension (see also Fig.~\ref{fig:spacetime})
\begin{eqnarray}
\mathcal{Z} &=& \int \mathcal{D} \psi (\tau) \mathcal{D} \phi_\alpha (x, \tau) \exp\left( - \mathcal{S}_\psi - \mathcal{S}_\phi \right) \nn
\mathcal{S}_\psi &=& \int d \tau \,
\psi^\dagger \left( \frac{\partial}{\partial \tau} + \lambda - \gamma_0 \phi_1 (x=0,\tau) \right) \psi  \nonumber \\
\mathcal{S}_{\phi} &=& \int d^d x \int d \tau \left[ \frac{1}{2} (\partial_\tau \phi_{\alpha})^2 +
\frac{1}{2} (\partial_x \phi_{\alpha})^2 + \frac{s}{2}\phi_{\alpha}^2 +
\frac{g_0}{4!} (\phi_{\alpha}^2)^2 \right]. \label{efermion}
\end{eqnarray}
Note that the fermion $\psi$ does not carry an O($N$) index, and we have chosen it to couple only to the $\alpha=1$ component of $\phi_\alpha$:
thus the fermion breaks the O($N$) symmetry of the Wilson-Fisher theory in its vicinity.
Here $\lambda$ is the energy of the dispersionless band (which will be implicitly renormalized), $s$ is the coupling used to tune the bosonic
sector across the Wilson-Fisher fixed point, $g_0$ is the relevant self-interaction of the bosons, and $\gamma_0$ is the relevant `Yukawa' coupling
between the fermion and bosons.
Our basic result will be that when the bosonic sector is at the quantum critical point in $d<3$, the fermion Green's function obeys at real frequencies $\omega$
\beq
G (\omega)  \sim \frac{1}{(\lambda - \omega)^{1-\eta_\psi}}. \label{Gpsi}
\eeq
So there is no fermion quasiparticle pole, and we instead have ``non-Fermi liquid'' behavior characterized by the universal
anomalous dimension $\eta_\psi$.

When we move away from the strict flat band limit and include a small dispersion for the fermions, then 
we expect \cite{kachru1,kachru2,qimp3,qimp3a} that Eq.~(\ref{Gpsi}) is only modified by a momentum dependence in the value of the threshold $\lambda$. 
Thus for a Fermi surface with the incipient quasiparticle/hole dispersion $\varepsilon(k)$ which vanishes on the Fermi surface, we will have
for the quasiparticle-remnant Green's function
\beq
G_{\rm qp} (k,\omega)  \sim \frac{1}{(\varepsilon (k) - \omega)^{1-\eta_\psi}} \quad , \quad \varepsilon(k) > 0,  \label{Gqp}
\eeq
and for the quasihole-remnant Green's function
\beq
G_{\rm qh} (k,\omega)  \sim - \frac{1}{(-\varepsilon (k) + \omega)^{1-\eta_\psi}} \quad , \quad \varepsilon(k) < 0. \label{Gqh}
\eeq
The Green's function in Eq.~(\ref{Gqp}) has a non-zero imaginary part for $\omega > \varepsilon (k)$, while that in Eq.~(\ref{Gqh}) has a non-zero imaginary part for $\omega < \varepsilon (k)$.

We will compute $\eta_\psi$ in an expansion in $\epsilon=3-d$. An unusual feature of our $\epsilon$-expansion is that the Yukawa coupling
$\gamma_0$ is of order unity at the fixed point which yields Eq.~(\ref{Gpsi}) (this is a significant difference from Ref.~\onlinecite{kachru2} where a different model is considered in which the fixed point is at small $\gamma_0$). 
This implies that our analysis must be carried out to 
{\em all orders\/} in $\gamma_0$, and we will show how this can be accomplished. The coupling $g_0$ is of order $\epsilon$ at the fixed point (as usual),
and so an expansion in powers of $g_0$ is permitted. Because of this novel structure in the $\epsilon$-expansion, we find that we have to evaluate
Feynman diagrams which include up to 4 loop momenta to obtain results even to first order in $\epsilon$; such a computation yields
\beq
\eta_\psi = \frac{(N+8)}{4 \pi^2} \left[ 1 + \frac{(1.68269 N^2 + 17.4231 N + 64.6922)}{(N+8)^2} \epsilon + \mathcal{O}(\epsilon^2) \right] \label{etaval}
\eeq
The exact value of the co-efficient of the $\epsilon$ term is given in Eq.~(\ref{mainres}), where 
the values of the numbers $\mathcal{C}_{1,2,3}$ are specified in Eqs.~(\ref{c12},\ref{c3}) in terms of digamma and zeta functions; for the case $N=1$ relevant to the Ising-nematic critical point, the co-efficient of the $\epsilon$ term is 1.0354.

Another notable feature of Eq.~(\ref{etaval}) is that $\eta_\psi$ does not vanish as $\epsilon \rightarrow 0$.
However, it is not the case that the problem in $\epsilon=0$ ({\em i.e.\/} in $d=3$) is characterized by the universal $\eta$ in Eq.~(\ref{etaval}).
The $\epsilon=0$ case will be briefly mentioned in the body of the paper, and it has a non-universal $\eta_\psi$ dependent upon
bare couplings. 
Thus the $\epsilon=0$ physics is different from the $\epsilon \rightarrow 0$ limit. This subtlety is related to the 
requirement noted above of having to compute results to all orders in $\gamma_0$. In a similar vein, note that the value of 
$\eta_\psi$ appears to diverge as $N \rightarrow \infty$ at fixed $\epsilon$ in Eq.~(\ref{etaval}). This is not expected to be correct,
and the divergence is expected to be
absent because the $\epsilon \rightarrow 0$ and $N \rightarrow \infty$ limits to not commute: Eq.~(\ref{etaval}) is only valid as $\epsilon \rightarrow 0$ at fixed $N$. We will consider other aspects of the $N \rightarrow \infty$ 
limit at fixed non-zero $\epsilon$ in Section~\ref{sec:largeN}, and find there that a large $N$ solution exists only for $\epsilon > 1/2$.

Our analysis of $\mathcal{Z}$ will be aided by its connection to an auxiliary problem in which the fermion $\psi$ is eternally
present at $x=0$; this connection is similar to that between the traditional X-ray edge and Kondo problems, and was
pointed out in Ref.~\onlinecite{ssimpurity} for a closely related problem. With the fermion present, the Yukawa
coupling in Eq.~(\ref{efermion}) becomes equivalent to a {\em local field\/} acting on the $\alpha=1$ component of
the order parameter along a defect line at $x=0$ and all $\tau$. So we are led to consider
\beq
\mathcal{Z}_d = \int \mathcal{D} \phi_\alpha (x, \tau) \exp \left( - 
\mathcal{S}_{\phi} - \gamma_0 \int d \tau
\phi_{1} (x=0, \tau) \right) \label{e1}
\eeq
This partition function $\mathcal{Z}_d$ is characterized by the same couplings $\gamma_0$ and $g_0$ as 
$\mathcal{Z}$, and they will have identical beta-functions in the two problems.\cite{ssimpurity} Indeed, the beta-functions
are easier to compute in the $\mathcal{Z}_d$ formulation, and we will exploit this feature. However, the exponent $\eta_\psi$ can only be computed in the $\mathcal{Z}$ formulation, which we have to use to compute the overlap between quantum states in which the fermion is present and absent.

The physics of the $\mathcal{Z}_d$ formulation is similar to that of numerous other analyses of defect lines in 
CFTs.\cite{qimp1,qimp2,qimp3,qimp3a,qimp4,kachru3,gaiotto1,gaiotto2}
In the $\mathcal{Z}_d$ formulation we can examine the behavior of the boson correlations as they approach the defect
line at $x=0$; in $\mathcal{Z}_h$, these would correspond to $\phi_\alpha$ correlations near the fermion long after it has been created. 
The coupling $\gamma_0$ flows to a fixed-point value, and so there is a strong local field that acts on $\phi_\alpha$ at $x=0$:
this suggest that in the operator product expansions the  
bulk $\phi_\alpha$ operator can be replaced by the constant unit operator near the defect line.\cite{qimp4}
In this situation we expect that \cite{qimp4}
\beq
 \left\langle \phi_{\alpha} (x, \tau) \right\rangle \sim \frac{\delta_{\alpha,1}}{x^{(d-1+\eta)/2}},
 \label{unit}
\eeq
where $\eta$ is bulk anomalous dimension of the Wilson-Fisher theory. We will find results consistent
with Eq.~(\ref{unit}) in 
an $\epsilon$ expansion computation in Section~\ref{sec:defect}, and a large $N$ computation in Section~\ref{sec:largeN}.

The outline of the remainder of the paper is as follows. Section~\ref{sec:defect} will present a computation of the beta functions
in the line defect model $\mathcal{Z}_d$. The $\epsilon$ expansion for the fermion anomalous dimension $\eta_\psi$ associated
with $\mathcal{Z}$ appears in Section~\ref{sec:fermion}. Finally, in Section~\ref{sec:largeN} we return to $\mathcal{Z}_d$
and analyze it in the large $N$ expansion in general $d$.

\section{Line defect in the Wilson-Fisher theory}
\label{sec:defect}

A number of earlier works have considered line defects in the 2+1 dimensional CFT described by the Wilson-Fisher fixed point.\cite{qimp1,troyerprl,ssimpurity,qimp3,qimp3a,gaiotto1,gaiotto2}
However, none of these works considered the case of interest to us here as described by $\mathcal{Z}_d$ in Eq.~(\ref{e1}): 
a local field acting at $x=0$ which locally breaks the O($N$)
symmetry of the bulk theory. 

Our analysis of $\mathcal{Z}_d$ begins by recalling the well-known\cite{bgz} renormalization of the bulk theory, which remains unmodified by
the presence of the impurity. We define renormalized fields and couplings by
\begin{equation}
\phi_{\alpha} = \sqrt{Z} \phi_{R \alpha}~~~~;~~~~g_0 =
\frac{\mu^{\epsilon} Z_4}{Z^2 S_{d+1}} g . \label{gren}
\end{equation}
Here $\mu$ is a renormalization momentum scale, and 
\beq
S_d = \frac{2}{\Gamma(d/2) (4 \pi)^{d/2}} 
\eeq
is a phase space factor.
The renormalization constants $Z$, $Z_4$ were computed long ago
\cite{bgz}; their values in the minimal subtraction scheme to
order $g^2$ are
\begin{equation}
Z = 1 - \frac{(N+2) g^2}{144 \epsilon} + \mathcal{O} (g^3)~~~;~~~Z_4 = 1 + \frac{(N+8)g}{6
\epsilon} + \left(\frac{(N+8)^2}{36 \epsilon^2} - \frac{(5N+22)}{36
\epsilon} \right)g^2 + \mathcal{O} (g^3). \label{e3a}
\end{equation}

\begin{figure}
\includegraphics[width=3.5in]{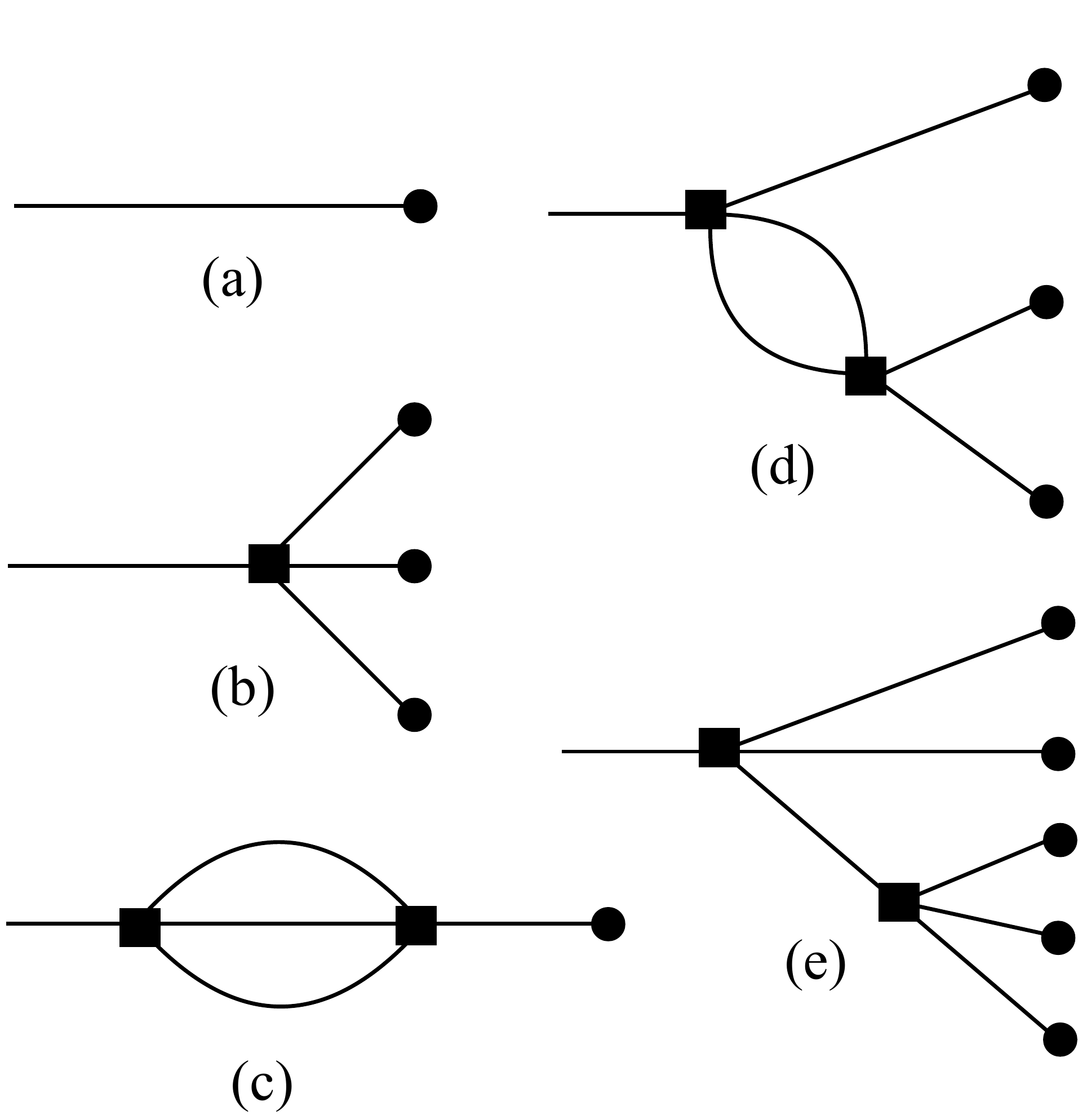} 
\caption{Feynman diagrams for $\langle \phi \rangle$ to order $g_0^2$. The full line is the bulk $\phi$ propagator under
$\mathcal{S}_\phi$, the filled square is the bulk coupling $g_0$, and the filled circle is the boundary coupling $\gamma_0$.
}
\label{fig:feyn1}
\end{figure}

We now compute the `boundary' renormalizations associated with the defect line. First we define
the renormalization
\begin{equation}
\gamma_0 = \frac{\mu^{\epsilon/2}
Z_{\gamma}}{\sqrt{Z \widetilde{S}_{d+1}}} \gamma. 
\label{gammaren}
\end{equation}
where $Z_\gamma$ is the new impurity renormalization, and the phase space factor $\widetilde{S}_{d+1}$ is defined 
below in Eq.~(\ref{phase}). 
To evaluate $Z_\gamma$, we first compute the expectation value of $\phi_1$ to second order in bare perturbation theory in $g_0$,
but to all orders in the boundary coupling $\gamma_0$. All Feynman diagrams to this order are shown in Fig.~\ref{fig:feyn1}.
These diagrams are most conveniently evaluated by going back-and-forth between propagators in real and momentum space. 
The bulk real space propagator for the $\phi$ field is
\bea
D_0 (x,\tau) &=& \int \frac{d^d k d \omega}{(2 \pi)^{d+1}} \frac{e^{- i (k x + \omega \tau)}}{\omega^2 + k^2} \nn
&=& \frac{\widetilde{S}_{d+1}}{(x^2 + \tau^2)^{(d-1)/2}}.
\eea
We also repeatedly use the Fourier transform (and its inverse)
\beq
\int d^d x \frac{e^{- i k x}}{x^a} = \frac{S_{d,a}}{k^{d-a}}
\eeq 
where
\beq
S_{d,a} \equiv
\frac{2^{d-a} \pi^{d/2} \Gamma((d-a)/2)}{\Gamma(a/2)} \quad, \quad \widetilde{S}_d = \frac{S_{d,2}}{(2 \pi)^d}
\label{phase}
\eeq
The diagrams in Fig.~\ref{fig:feyn1} yield for the expectation value of the renormalized field
\beq
\left\langle \phi_{R1} (k) \right\rangle = \frac{1}{\sqrt{Z}} \left[ {\rm (a)}+{\rm (b)}+{\rm (c)}+{\rm (d)}+{\rm (e)} \right] \label{phiR}
\eeq
where
\bea
{\rm (a)} &=& \frac{\gamma_0}{k^2} \nn
{\rm (b)} &=& - \frac{g_0 \gamma_0^3}{6} \, \frac{\widetilde{S}_{d}^3 S_{d,3d-6}} {k^{8-2d}} \nn
{\rm (c)} &=& \frac{(N+2) g_0^2 \gamma_0}{18} \, \frac{\widetilde{S}_{d+1}^3 S_{d+1,3d-3}} {k^{8-2d}} \nn
{\rm (d)} &=& \frac{(N+8) g_0^2 \gamma_0^3}{36} \, \frac{\widetilde{S}_{d+1}^2 \widetilde{S}_d^3 S_{d+1,2d-2} S_{d,2d-4} S_{d,7-2d} 
S_{d,4d-9}}{(2 \pi)^d k^{11-3d}} \nn
{\rm (e)} &=& \frac{g_0^2 \gamma_0^5}{12} \, \frac{\widetilde{S}_d^5 S_{d,3d-6} S_{d,8-2d} S_{d,5d-12}}{(2 \pi)^d k^{14-4d}}
\eea
Now we express Eq.~(\ref{phiR}) in terms of the renormalized couplings in Eqs.~(\ref{gren},\ref{gammaren}), expand the resulting expression in powers of $g$
(but not $\gamma$), and demand that all poles in $\epsilon$ cancel at each order in $g$. This yields the following value of $Z_\gamma$ in the minimal
subtraction scheme
\beq
Z_\gamma = 1 + \frac{\pi^2 g \gamma^2}{6 \epsilon} + 
g^2 \left( 
\frac{\pi^2 \gamma^2 (2N + 16 + 9 \pi^2 \gamma^2)}{216 \epsilon^2} + \frac{\mathcal{C}_1 \gamma^2 + \mathcal{C}_2 \gamma^4}{\epsilon}
\right) + 
\mathcal{O} (g^3) 
\label{Zgamma}
\eeq
where the numerical constants $\mathcal{C}_{1,2}$ are
\bea
\mathcal{C}_1 &=& \frac{\pi^2}{216} \left( - 3- 2 \gamma_E + \ln(4) - \psi(1/2) + \psi(3/2) \right)
\nn
\mathcal{C}_2 &=& \frac{\pi^4}{24} \left( \psi (1/2) - \psi (3/2) \right), \label{c12}
\eea
with $\gamma_E$ the Euler-Mascheroni constant and $\psi$ the digamma function.

With all renormalization constants determined, we can now compute the beta-functions for the couplings $g$ and $\gamma$
\bea
\beta (g) &=& - \epsilon g + \frac{(N+8)}{6} g^2 - \frac{(3N+14)}{12} g^3  + \mathcal{O} (g^4) \nn
\beta (\gamma) &=& - \frac{\epsilon}{2} \gamma  + \frac{\pi^2}{3}  \gamma^3 g + 
\left( \frac{(N+2)}{144} \gamma + 3 (N+8) \mathcal{C}_1 \gamma^3 + 4 \mathcal{C}_2 \gamma^5 \right) g^2
+
\mathcal{O} (g^3) \label{betafns}
\eea
All poles in $\epsilon$ cancel in this computation, verifying the renormalizability of the theory.
Note that at $g=0$, the flow of $\gamma$ is just given by its naive scaling dimension: 
in the context of the fermion theory in Eq.~(\ref{efermion}), this is a consequence of an
exact cancellation between fermion 
self-energy and vertex corrections which is described in Appendix~\ref{app:vertex} (Refs.~\onlinecite{kachru1,kachru2} examined
models which did not have this cancellation). However, once the bulk interactions
associated with the scalar field are included, there is a non-trivial flow of $\gamma$.

These beta-functions have the infrared attractive fixed point
\bea
g^\ast &=& \frac{6}{(N+8)} \epsilon + \frac{18 (3N+14)}{(N+8)^3} \epsilon^2 + \mathcal{O}(\epsilon^3) \nn
\gamma^{\ast 2} &=& \frac{(N+8)}{4 \pi^2}\left[1 - \left( \frac{19N+86}{2(N+8)^2} + \frac{54 \mathcal{C}_1}{\pi^2} +  \frac{18 \mathcal{C}_2}{\pi^4}\right) \epsilon
+ \mathcal{O} (\epsilon^2) \right] \label{fp}
\eea
Note that $\gamma^\ast$ remains finite as $\epsilon \rightarrow 0$, as we emphasized in Section~\ref{sec:intro}. However, precisely
in $\epsilon=0$, analysis of Eq.~(\ref{betafns}) shows that $g$ approaches the fixed point $g^\ast=0$, while $\gamma$ does 
{\em not\/} approach a fixed point. Consequently, the $\epsilon \rightarrow 0$ limit is distinct from the $\epsilon=0$ case.

We can now reinsert these fixed point values into our expansion in Eq.~(\ref{phiR}) for $\langle \phi \rangle$; at the critical point we
find that the results at order $\epsilon^2$ are compatible with Eq.~(\ref{unit}), which implies the following expression
for the expectation value in momentum space
\beq
\left\langle \phi_{R1} (k) \right\rangle = \mathcal{N} \frac{2 \pi \gamma^\ast}{k^{2-\epsilon/2}}  \left(\frac{k}{\mu} \right) ^{\eta/2} ,
\eeq
where
\beq
\eta = \frac{(N+2)}{2(N+8)^2} \epsilon^2 + \mathcal{O}(\epsilon^3),
\eeq
is the bulk scaling dimension of $\phi$. Eq.~(\ref{phiR}) yields that 
\beq
\mathcal{N} =  
1 - 1.0844693 \epsilon + \left( 0.579032 - 0.235508 \frac{(N+2)}{(N+8)^2} \right) \epsilon^2 + \mathcal{O}(\epsilon^3).
\eeq

\section{Fermion anomalous dimension}
\label{sec:fermion}

We now examine the theory $\mathcal{Z}$ in Eq.~(\ref{efermion}).

In the conventional renormalization scheme, the renormalizations of $\mathcal{S}_\phi$ remain the same as those of the bulk theory
as described in Section~\ref{sec:defect}, while
for the fermion sector we introduce the wavefunction renormalization
\beq
\psi = \sqrt{Z_{\psi}} \psi_{R}
\eeq
and the renormalization of the ``Yukawa'' coupling
\begin{equation}
\gamma_0 = \frac{\mu^{\epsilon/2}
\widetilde{Z}_{\gamma}}{Z_{\psi} \sqrt{Z \widetilde{S}_{d+1}}} \gamma. 
\label{gammarenh}
\end{equation}
Note that this renormalization scheme differs from that in Eq.~(\ref{gammaren}). However, we expect the RG flow of the underlying 
coupling to be the same in the two theories, and so we conclude that \cite{ssimpurity}
\beq
\widetilde{Z}_\gamma = Z_{\psi} Z_\gamma . \label{gammah}
\eeq
We explicitly verify this identity at low orders in Appendix~\ref{app:vertex}.

We are primarily interested here in the wavefunction renormalization of the fermion, $Z_{\psi}$.
We proceed by introducing a `gauge-transformed' fermion field $\overline{\psi}$
\beq
\psi (\tau) = 
\overline{\psi} (\tau) \exp \left( \gamma_0 \int_0^\tau d \tau_1 \phi_1 (x=0,\tau_1) \right) \label{gauget}
\eeq
Now $\overline{\psi}$ is a free fermion, and so correlators of $\psi$ can be evaluated using the exact expression
\beq
G(\tau) = G_0 (\tau) \left\langle
\exp \left( \gamma_0 \int_0^\tau d \tau_1 \phi_1 (x=0,\tau_1) \right) \right\rangle_{\mathcal{S}_\phi}
 \label{G0}
\eeq
where $G_0$ is the free fermion correlator
\beq
G_0 (\tau) = e^{-\lambda \tau} \theta(\tau).
\eeq
The two-point
correlators of $\phi_1 (x=0, \tau)$ given by
\bea
D_0 (x,\tau) &=& \int \frac{d^d k d \omega}{(2 \pi)^{d+1}} \frac{e^{- i (k x + \omega \tau)}}{\omega^2 + k^2} \nn
&=& \frac{\widetilde{S}_{d+1}}{(x^2 + \tau^2)^{(d-1)/2}}.
\eea
So at order $g^0$ we have for the fermion Green's function
\beq
G(\tau) = G_0 (\tau) \exp \left( \frac{\gamma_0^2}{2} \int_0^\tau d \tau_1 \int_0^\tau d \tau_2 \, D_0 (0, \tau_1 - \tau_2) \right) \label{G1}
\eeq

It is physically more transparent to momentarily impose a short-distance cutoff, $a$,
in $D_0$
\beq
D_0 (x, \tau) \rightarrow \widetilde{S}_{d+1} \, \frac{(1 - e^{-(x^2+\tau^2)/a^2})}{(x^2 + \tau^2)^{(d-1)/2}},
\eeq
and evaluate Eq.~(\ref{G1}) in $d=3$ for large $\tau>0$
\beq
G(\tau) = G_0 (\tau) \exp \left( \frac{\gamma_0^2 \widetilde{S}_4}{2} \left[ \frac{2 \sqrt{\pi} \tau}{a}  - 2 \ln \left( \frac{\tau}{a} \right)+ \ldots \right] \right).
\eeq
The leading term in the exponential is absorbed into a renormalization of the fermion energy $\lambda$, while the second term yields the anomalous
dimension of the fermion
\beq
\eta_\psi  = \gamma^2 + \mathcal{O} (g). \label{r5a}
\eeq
As $\gamma$ does not approach a fixed point value for $\epsilon=0$, there is no universal anomalous dimension in $d=3$. However, for
$\epsilon >0$, there is a fixed point as we noted below Eq.~(\ref{betafns}), and $\eta_\psi$ is therefore universal.

Determination of the $\mathcal{O}(g)$ term will be carried out using the dimensional regularization method.
In this method, Eq.~(\ref{G1}) yields
\bea
\frac{G(\tau)}{G_0 (\tau)} &=&  \exp \left(  \frac{\gamma_0^2 \widetilde{S}_{d+1}}{2} \int_0^\tau d \tau_1 \int_0^\tau d \tau_2 \,
\frac{1}{|\tau_1-\tau_2|^{2-\epsilon}} \right) \nn
&=&\exp \left( - \frac{\gamma^2 Z_\gamma^2}{Z} \frac{(\tau\mu)^{\epsilon}}{\epsilon (1-\epsilon)}  \right). \label{tau1}
\eea
Now demanding that poles cancel in the Green's function of the renormalized field $\psi_R$ we obtain
\beq
Z_{\psi} = \exp \left( - \frac{\gamma^2}{\epsilon} \right)  + \mathcal{O} (g),
\label{Z1}
\eeq
and this leads to an anomalous dimension in agreement with Eq.~(\ref{r5a})
\begin{equation}
\eta_\psi = \beta (\gamma) \frac{d \ln Z_{\psi}}{d \gamma} = 
\gamma^2 + \mathcal{O} (g). \label{r5}
\end{equation}

At next order in $g$, evaluation of Eq.~(\ref{G0}) shows that 
\bea
G(\tau) &=& G_0 (\tau) \exp \left( \frac{\gamma_0^2}{2} \int_0^\tau d \tau_1 \int_0^\tau d \tau_2 \, D_0 (0, \tau_1 - \tau_2) \right) \nn
&~&~~ \times \left\{ 1 - \frac{g_0 \gamma_0^4}{24}\int d^d x \int_{-\infty}^{\infty} d \tau_0 \left[ \int_0^\tau d \tau_3  \, D_0 (x, \tau_3 - \tau_0) \right]^4  \right\} + \mathcal{O}(g^2), \label{3loop}
\eea
where we have dropped `tadpole' contributions which vanish in dimensional regularization.
The order $g$ term above is computed in
Appendix~\ref{app:3loop}, and from this we obtain the order $g$ correction to Eq.~(\ref{Z1}): this requires a 3-loop computation and leads to 
\beq
Z_{\psi} = \exp \left(- \frac{\gamma^2}{\epsilon} \right) \left[ 1 -  g \gamma^4 \left( \frac{2\pi^2}{9 \epsilon^2} + \frac{\mathcal{C}_3}{\epsilon} \right) + \mathcal{O} (g^2) \right] \label{Zhres}
\eeq
where from Eq.~(\ref{Cres}) we have ($\psi$ is the digamma function)
\bea
\mathcal{C}_3 &=& \frac{1}{18} \left[ 6 \zeta (3)+\pi ^2 \left(1+\ln (64)+3 \psi
   \left({1}/{2}\right)-\psi
   \left({3}/{2}\right)\right) \right]
    \label{c3}
%   \nn
%&=& -0.020492588211 \ldots   
\eea
We also computed $Z_{\psi}$ in a more conventional Dyson formulation in frequency space: the computations of the
frequency dependent self energy of the fermion is described in Appendix~\ref{app:selfen}, and yields a result for
$Z_{\psi}$ in perfect agreement with Eq.~(\ref{Zhres}).

Now the fermion anomalous dimension is
\bea
\eta_\psi &=& \beta (\gamma) \frac{d \ln Z_{\psi}}{d \gamma}  + \beta(g)  \frac{d \ln Z_{\psi}}{d g} \nn
&=&  \gamma^2  + 3 \mathcal{C}_3 \, g \gamma^4 + \mathcal{O} (g^2). \label{rf}
\eea
Note that the poles in $\epsilon$ have all cancelled in Eq.~(\ref{rf}): this is a highly non-trivial check of our computation.
We now insert the fixed-point values of the couplings in Eq.~(\ref{fp}) and obtain our main result
\beq
\eta_\psi = \frac{(N+8)}{4 \pi^2} \left( 1 - \left( 
\frac{19N+86}{2 (N+8)^2} + \frac{9(12 \pi^2 \mathcal{C}_1 + 4 \mathcal{C}_2 - \pi^2 \mathcal{C}_3)}{2 \pi^4} 
\right) \epsilon + \mathcal{O}(\epsilon^2) \right); \label{mainres}
\eeq
the values of the numbers $\mathcal{C}_{1,2,3}$ above are specified in Eqs.~(\ref{c12},\ref{c3}) in terms of digamma and zeta functions.

\section{Large $N$ analysis of defect line}
\label{sec:largeN}

This section returns to the defect line model in Eq.~(\ref{e1}) and analyzes it in the limit of large $N$ for general $d$.

We formulate the large $N$ limit using a theory with a fixed-length constraint $\sum_{\alpha=1}^N \phi_\alpha^2 = \mbox{constant}$.
This constraint is implemented by a Lagrange multiplier $\lambda$. After suitable rescalings of fields and couplings for a 
useful large $N$ limit, the action for $\mathcal{Z}_d$ in Eq.~(\ref{e1}) is modified
to
\beq
\mathcal{S}_d =  \int d^d x \int d \tau \frac{1}{2 g} \left[  (\partial_\tau \phi_{\alpha})^2 +
 (\partial_x \phi_{\alpha})^2 + i \lambda (\phi_\alpha^2 - N)
 \right] - \gamma_0 \sqrt{N} \int d \tau
\phi_{1} (x=0, \tau) .
\eeq
We are now using the coupling constant $g$ to tune the bulk theory across its quantum critical point.

We now parameterize
\beq
\phi_{\alpha} = (\sqrt{N} \sigma, \pi_1 , \pi_2 , \ldots, \pi_{N-1} )
\eeq
and integrate out the $\pi$ fields. Then action becomes
\bea
\mathcal{S}_d &=&  \int d^d x \int d \tau \frac{N}{2 g} \left[  (\partial_\tau \sigma)^2 +
 (\partial_x \sigma)^2 + i \lambda (\sigma^2 - 1)
 \right] - \gamma_0 N \int d \tau
\sigma (x=0, \tau) \nn 
&~&~~~+ \frac{N-1}{2} \mbox{Tr} \ln \left[ - \partial_\tau^2 - \partial_x^2 + i \lambda \right] \label{Sd}
\eea
So in the large $N$ limit involves determination of the saddle point of Eq.~(\ref{Sd}) with
respect to the space-dependent fields $\sigma (x)$ and $\lambda (x)$.

In the absence of the external field, $\gamma_0 = 0$, the critical point is at\cite{csy} $g=g_c$, where
\beq
\frac{1}{g_c} = \int \frac{d \omega d^d k}{(2 \pi)^d} \frac{1}{\omega^2 + k^2}
\eeq
At the critical point, the saddle point value $i \lambda = 0$.

In the presence of a field, we expect a saddle point with $i \lambda = \Delta^2 (x)$ 
and $\sigma = \sigma (x)$, with 
$\Delta (x), \sigma(x) \rightarrow 0$ as $|x| \rightarrow \infty$. 
The saddle-point equations determining these functions are
\beq
\left [-\nabla_x^2 + \Delta^2(x)\right ] \sigma(x) = \gamma_0 g_c\delta^d(x) \label{saddle1}
\eeq
and
\beq
\sigma^2(x) + g_c G(x,\tau; x,\tau) = 1 \label{saddle2}
\eeq
where $G$ is the Green's function obeying
\begin{equation}
 \left [-\nabla_x^2 -\partial_\tau^2 + \Delta^2(x)\right ] G(x,\tau; x',\tau') = \delta^d(x - x')\delta(\tau - \tau') \,.
\end{equation} 
It is useful to write Eq.~(\ref{saddle2}) as
\begin{equation}
  \sigma^2(x) + g_c \left[ G(x,\tau; x,\tau) - G_0(x,\tau; x,\tau)\right] = 0\,,
\end{equation} 
where $G_0 = G|_{\Delta\equiv 0} = 1/g_c$. For $d < 3$ the difference $G - G_0$ is ultraviolet (u.v.) finite. 
There is however a u.v. divergence associated with the Dirac delta in the first equation, which will become manifest later.

Now we introduce the orthonormal and complete set of eigenfunctions
\begin{equation}
  \left [-\nabla_x^2 + \Delta^2(x)\right ] \psi_n(x) = q_n^2 \psi_n(x)\,,
\end{equation} 
and we express $\sigma$ and $G$ in terms of these
\begin{align}\label{eq:expansion1}
 &G(x, \tau; x', \tau')  = \int \frac{\dd \omega}{2\pi} \sum_{n}  \frac{1}{q_n^2 + \omega^2} \psi_n(x) \psi^\star_n(x')e^{\ii \omega (\tau - \tau')}\,,\\
 &\sigma(x) = \gamma_0 g_c \sum_{n} \frac{1}{q_n^2} \psi_n(0) \psi^\star_n(x')\,.
\end{align} 

Guided by rotational and scale invariance, we assume that
\begin{equation}
 \Delta^2(x) = \frac{v}{x^2}\,.
\end{equation} 

It is advantageous to expand the eigenfunctions $\psi_n$ over the orthonormal spherical harmonics $Y_{\ell m}$ of the $d-1$ sphere:
\begin{equation}
 \psi_{n}(r, \Omega) = \psi_{q\ell}(r) Y_{\ell m}(\Omega)\,,
\end{equation} 
where the radial wavefunction $\psi_{q\ell}$ satisfies the eigenvalue equation
\begin{equation}
 \left[-\partial_r^2 - \frac{d-1}{r}\partial_r + \frac{\ell(\ell + d - 2) + v}{r^2}\right] \psi_{q\ell}(r) = q^2\psi_{q\ell}(r)\,.
\end{equation} 

The regular solution can be written in terms of Bessel functions:
\begin{align}\label{eq:nuell}
 &\psi_{q\ell} = r^{\frac{1-d}{2}} \sqrt{q r} J_{\nu_\ell}(q r)
 & \nu_\ell = \sqrt{\ell(\ell + d - 2) + v + (d/2-1)^2}\,,
\end{align} 
and it is normalized
\begin{equation}
 \int_{0}^{\infty} \dd r\ r^{d - 1} \psi_{q\ell}(r)\psi_{q'\ell}(r) = \delta(q- q')\,.
\end{equation} 

Substituting in (\ref{eq:expansion1}) we have
\begin{align}
 &G(r, \Omega, \tau; r, \Omega, 0)  = \frac{1}{A_{d-1}}\int \frac{\dd \omega}{2\pi} \int_{0}^\infty \dd q \sum_{\ell} {\rm deg}_{d\ell}\  \frac{1}{q^2 + \omega^2}[\psi_{q\ell}(r)]^2 e^{\ii \omega \tau}\,,\\
 &\sigma(r) = \frac{\gamma_0 g_c}{A_{d-1}} \int_{0}^\infty \dd q \sum_{\ell} {\rm deg}_{d\ell}\ \frac{1}{q^2}\psi_{q\ell}(\epsilon)\psi_{q\ell}(r)\,,
\end{align}  
where $\tau$ and $\epsilon$ are u.v. regulators, $A_{d}$ is the area of the $d$-sphere, and ${\rm deg}_{d\ell}$ is the degeneracy of the eigenspace of $L^2$ with eigenvalue $\ell(\ell + d - 2)$. In obtaining this result we used the identity
\begin{align}
 \sum_m \left|Y_{\ell m}(\Omega)\right|^2 = \frac{\mathrm{deg}_{d\ell}}{A_{d-1}}\,, && 
\begin{tabular}{ccc}
 $d$ & $A_{d-1}$ & $\mathrm{deg}_{d\ell}$ \\
\hline
 2 & $2\pi$ & 1 if $\ell = 0$, 2 otherwise\\
 3 & $4\pi$ & $2\ell + 1$\\
 d & $\displaystyle \frac{d \pi^{\frac{d}{2}}}{\Gamma\left(\frac{d}{2}+1\right)}$ & $\binom{d + \ell -1}{d - 1} - \binom{d + \ell -3}{d - 1}$
\end{tabular}
\end{align} 

The integrals over $\omega$ and $q$ can be done analytically
\begin{align}
 &G(r, \Omega, \tau; r, \Omega, 0) = \frac{1}{A_{d-1}r^{d-1}} \sum_{\ell}{\rm deg}_{d\ell}\   Q_\ell\left(\frac{\tau^2}{4r^2}\right)\,,\\
 &\sigma(r) = \frac{\gamma_0 g_c}{2A_{d-1}r^{d-2}} \sum_{\ell} {\rm deg}_{d\ell}\ \frac{1}{\nu_{\ell}}\left(\frac{\epsilon}{r}\right)^{\nu_{\ell}}\,,
\end{align}
where
\begin{align}
  Q_\ell(z) &= \frac{1}{2\sqrt{\pi}} \frac{\Gamma\left(\nu + \frac{1}{2}\right)}{\Gamma\left(\nu + 1\right)} (4z)^{-\nu-\frac{1}{2}} F\left(\nu + \frac{1}{2}, \nu + \frac{1}{2}, 2\nu + 1;\ - \frac{1}{z}\right) \\
  &= \frac{1}{16 \pi} - \frac{1}{4\pi} \ln z - \frac{1}{2\pi} H_{\nu - \frac{1}{2}} + \mathcal{O} (z)\,,
\end{align} 
where $H_{n}$ is the harmonic number. The u.v. and infrared (i.r.) divergent term $\log z$ cancels when taking the difference between $G$ and $G_0$. Then it is safe to take the limit $\tau \to 0$, and we have
\begin{align}
 &G(r, \Omega, 0; r, \Omega, 0) - G_0(r, \Omega, 0; r, \Omega, 0) = -\frac{W_d}{2\pi  A_{d-1} r^{d-1}}\,, &&
W_d=  \sum_{\ell}{\rm deg}_{d\ell} \left[H_{\nu_\ell - \frac{1}{2}} - H_{\bar\nu_\ell - \frac{1}{2}}\right]\,,
\end{align}
where $\bar \nu_\ell$ is given by (\ref{eq:nuell}) with $v = 0$. The sum over $\ell$ is convergent for $d < 3$ and it is positive if $v > 0$.

For what concerns $\sigma$, the dominant contribution as $\epsilon \to 0$ is given by $\ell  = 0$
\begin{equation}
 \sigma(r) = \frac{\gamma_0 g_c}{2\nu_{0}A_{d-1}r^{d-2}} \left(\frac{\epsilon}{r}\right)^{\nu_{0}}\,.
\end{equation}
It is now apparent that the u.v. regulator $\epsilon$ can be adsorbed in a redefinition of $\gamma_0$.

Both $\sigma$ and $G - G_0$ have a power law dependence on $r$. A solution to the saddle point equations is possible only if the two power laws match. This fixes the coefficient $v$:
\begin{align}
 &\nu_0 = \frac{3 - d}{2} &&\text{i.e.} && v = \frac{5 - 2d}{4}\,.
\end{align}

For consistency we need $v > 0$ and hence $d < 5/2$. The fixed point bare coupling $\gamma_0$ is given by
\begin{equation}
 \gamma_0^2 = \frac{(3-d)^2 A_{d-1}W_d}{2\pi\epsilon^{3-d}g_c}\,,
\end{equation} 
and we have
\begin{align}
 &\sigma = \sqrt{\frac{W_d g_c}{2\pi A_{d-1}}} \frac{1}{r^{\frac{d-1}{2}}}\,,
 &&\Delta^2(r) = \frac{5-2d}{4}\frac{1}{r^2}\,.
\end{align} 
So our final result for $\sigma (r)$ is consistent with Eq.~(\ref{unit}) because the bosonic $\eta=0$
in the present large $N$ limit. 

It is also interesting to note that the large $N$ limit provides a consistent
scaling solution only for $d<5/2$. So evidently, there is no solution when both $\epsilon=3-d$ and $1/N$ are small:
this feature is consonant with our earlier observation that the $\epsilon\rightarrow 0$ and $N\rightarrow \infty$
limits do not commute.

\section{Conclusions}
\label{sec:conc}

The main potential applicability of the present theory is to the Ising-nematic quantum critical point of metals.\cite{kachru1,kachru2,raghu}
For suitable microscopic parameters, there can be an extended intermediate regime where the Landau damping of the bosonic order
parameter can be ignored, and the boson correlations have dynamic critical exponent $z=1$. In this regime, if the Fermi velocity $v_F$ scales
to zero, then the problem of determining the fermion spectrum reduces to that considered in the present paper.

The flow of $v_F$ to small values in this intermediate regime appears in a one-loop renormalization group analysis.\cite{kachru2}
An important direction for future research is examine the flow of $v_F$ beyond the one-loop level.

\acknowledgments
We thank E.~Berg, A.~L.~Fitzpatrick, E.~Fradkin, D.~Gaiotto, S.~Kachru, J.~Kaplan, S.~Kivelson, M.~Metlitski and S.~Raghu for useful discussions.
This research
was supported by the NSF under Grant DMR-1103860, and the Templeton foundation.
Research at Perimeter Institute is supported by the
Government of Canada through Industry Canada and by the Province of
Ontario through the Ministry of Research and Innovation.

\appendix

\section{Vertex renormalization} 
\label{app:vertex}

This appendix will compute the leading renormalizations of the Yukawa vertex, $\widetilde{Z}_\gamma$, 
and verify the identity in Eq.~(\ref{gammah}).

\begin{figure}
\includegraphics[width=3.5in]{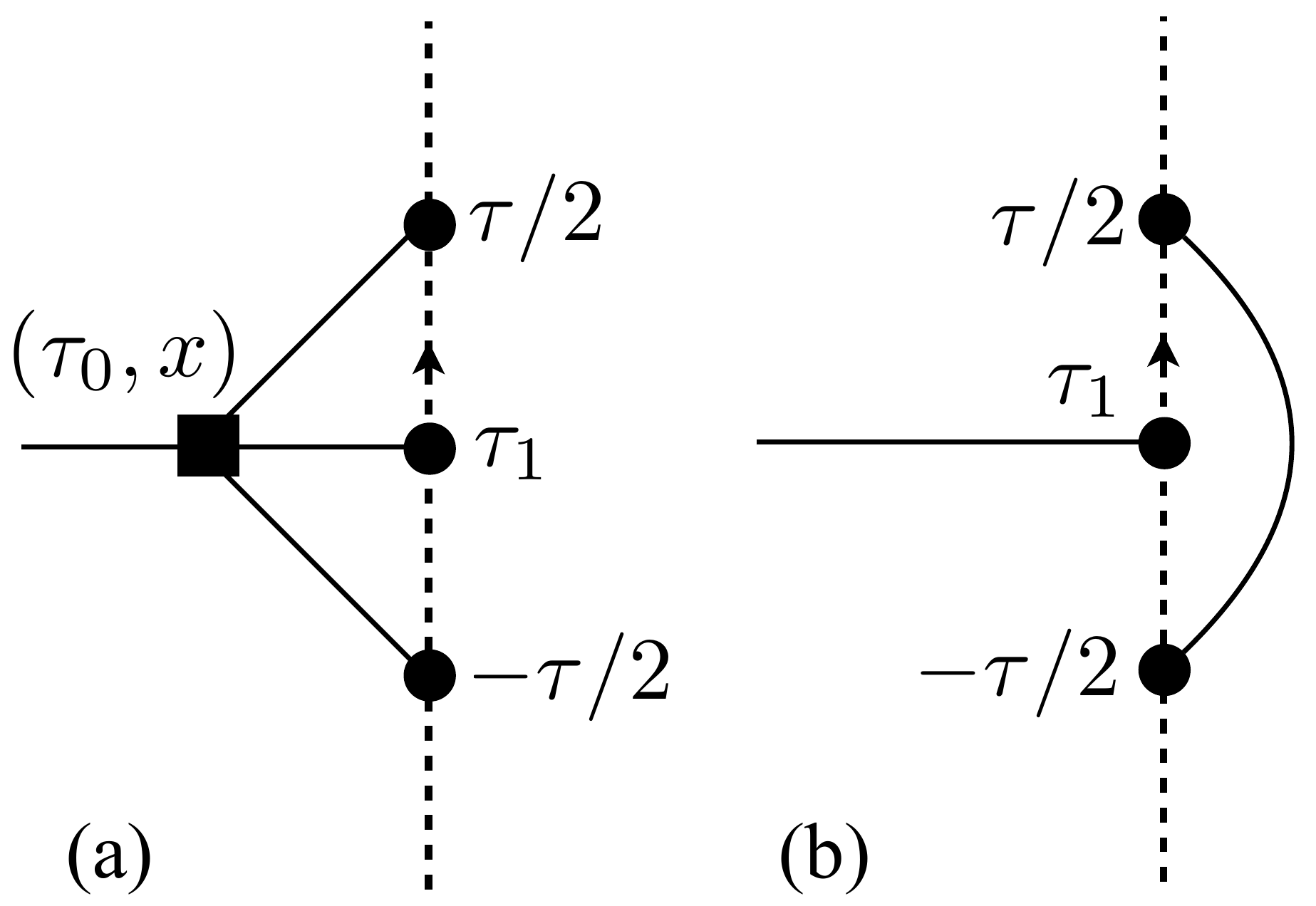} 
\caption{Feynman diagrams for the vertex renormalization. The dashed line is the fermion propagator
}
\label{fig:feyn2}
\end{figure}

The needed Feynman diagrams are shown in Fig.~\ref{fig:feyn2}, and they will be evaluated in real space and time.

From the diagram in Fig.~\ref{fig:feyn2}(a),
the vertex renormalization factor is
\bea
V_a (\tau)  &=& - \gamma_0^2 g_0 \widetilde{S}_{d+1}^3 \int d^d x \int_{-\infty}^{\infty} d \tau_0 
\frac{1}{\{[x^2 + (\tau_0 + \tau/2)^2][x^2 + (\tau_0 - \tau/2)^2]\}^{(d-1)/2}} \nn
&~&~~~~~~~~~~~~~~~~~~~~~~~~~~~~~~~~~~\times \left[ \int_{-\tau/2}^{\tau/2}  d \tau_1 
\frac{1}{[x^2 + (\tau_0 - \tau_1)^2]^{(d-1)/2}} \right] \nn
&=& - \frac{\mu^{2\epsilon}}{\tau^{1 - 2 \epsilon}} \, \frac{\gamma^2 g \widetilde{S}_{d+1}^2}{S_{d+1}} \int d^d x \int_{-\infty}^{\infty} d \tau_0 
\frac{1}{\{[x^2 + (\tau_0 + 1/2)^2][x^2 + (\tau_0 - 1/2)^2]\}^{(d-1)/2}} \nn
&~&~~~~~~~~~~~~~~~~~~~~~~~~~~~~~~~~~~\times \left[ \int_{-1/2}^{1/2}  d \tau_1 
\frac{1}{[x^2 + (\tau_0 - \tau_1)^2]^{(d-1)/2}} \right] \label{V1}
\eea
From the $\sim 1/\tau$ behavior of $V_a(\tau)$ at small $\epsilon$, we see that we will obtain a pole in $\epsilon$
in its Fourier transform $V_a(\omega)$. So, at leading order in $\epsilon$ we may evaluate all other terms at $\epsilon=0$, and obtain
\bea
V_a(\tau) &=& - \frac{\mu^{2 \epsilon}}{\tau^{1 - 2 \epsilon}} \, \frac{\gamma^2 g}{2 \pi^2} \int d^3 x \int_{-\infty}^{\infty} d \tau_0 
\frac{1}{[x^2 + (\tau_0 + 1/2)^2][x^2 + (\tau_0 - 1/2)^2]} \nn
&~&~~~~~~~~~~~~~~~~~~~~~~~~~~~~~~~~~~\times \left[ \int_{-1/2}^{1/2}  d \tau_1 
\frac{1}{[x^2 + (\tau_0 - \tau_1)^2]} \right] \nn
&=& - \frac{\mu^{2 \epsilon}}{\tau^{1 - 2 \epsilon}} \, \frac{\pi^2 \gamma^2 g}{3}
\eea
So after a Fourier transform
\beq
V_a(\omega) = (\mu/\omega)^{2 \epsilon} \left[ - \frac{\pi^2 g \gamma^2}{6 \epsilon} + \ldots \right]. \label{Varen}
\eeq
Similarly, from the diagram in Fig.~\ref{fig:feyn2}(b), the vertex renormalization is
\bea
V_b (\tau) &=& \gamma_0^2 \frac{\widetilde{S}_{d+1}}{\tau^{d-1}} \int_{-\tau/2}^{\tau/2} d \tau_1 \nn
&=& \frac{\mu^\epsilon}{\tau^{1-\epsilon}} \gamma^2.
\eea 
So the Fourier transform is
\beq
V_b (\omega) = (\mu/\omega)^{\epsilon} \left[ \frac{\gamma^2}{\epsilon} + \ldots \right]. \label{Vbren}
\eeq

Combining Eqs.~(\ref{Varen}) and (\ref{Vbren}), we obtain the vertex renormalization to order $g$ and $\gamma^2$
\beq
\widetilde{Z}_\gamma = 1 - \frac{\gamma^2}{\epsilon} + \frac{\pi^2 \gamma^2 g}{6 \epsilon}  + \ldots .
\eeq
This is in agreement with Eqs.~(\ref{Zgamma}), (\ref{gammah}) and (\ref{Z1}).
Notice that the vertex renormalization $V_b$ exactly cancels with the wavefunction renormalization in $Z_\psi$ at this order.
This is linked to our ability to solve the problem via the `gauge' transformation in Eq.~(\ref{gauget}).

\section{3-loop integral} 
\label{app:3loop}

This appendix will examine the following integral obtained from the $\mathcal{O}(g)$ term in Eq.~(\ref{3loop})
\bea
24 \, \mathcal{I} (\epsilon)  &\equiv& \frac{1}{S_{d+1} \widetilde{S}_{d+1}^2} \int d^d x \int_{-\infty}^{\infty} d \tau_0 \left[ \int_{-\tau/2}^{\tau/2} d \tau_3  \, D_0 (x, \tau_3 - \tau_0) \right]^4 \nn
&=& \tau^{3\epsilon} \, \frac{2 (2 \pi)^d S_d \widetilde{S}_{d+1}^2}{S_{d+1} }\int_0^\infty x^{2-\epsilon} dx \int_{0}^{\infty} d \tau_0 \left[ \int_{-1/2}^{1/2} d \tau_3  \, 
\frac{1}{(x^2 + (\tau_3 - \tau_0)^2)^{1-\epsilon/2}} \right]^4 \nn
&=& \tau^{3 \epsilon} \mathcal{A}_\epsilon
\int_0^\infty  x^{-2+3\epsilon} dx \, \Pi(x) , \label{IG}
\eea
where
\beq
\mathcal{A}_\epsilon \equiv \frac{2^{1-\epsilon}\, \Gamma^2 (1-\epsilon/2) \Gamma(2-\epsilon/2)}{\sqrt{\pi} \Gamma(3/2- \epsilon/2)} ,
\eeq
and
\beq
\Pi(x) = \int_{0}^{\infty} d \tau_0 \left[ \Phi (x, \tau_0) \right]^4 ,
\eeq
with 
\beq
\Phi (x, \tau_0) =  
 \frac{(1-2 \tau_0)}{2x} {}_2 F_1 \left( \frac{1}{2}, 1 - \frac{\epsilon}{2}, \frac{3}{2}, - \frac{(1-2 \tau_0)^2}{4 x^2} \right)
+ \frac{(1+2 \tau_0)}{2x} {}_2 F_1 \left( \frac{1}{2}, 1 - \frac{\epsilon}{2}, \frac{3}{2}, - \frac{(1+2 \tau_0)^2}{4 x^2} \right) 
\eeq
We can now write
\beq
\Pi (x) = x \int_{-1/(2x)}^{\infty} d \sigma \Biggl[ \sigma \phi_\epsilon (\sigma) - \left( \frac{1}{x} + \sigma \right) \phi_\epsilon \left( \frac{1}{x} + \sigma \right) \Biggr]^4
\eeq
where 
we introduced the variable $\sigma = (2 \tau_0 - 1)/(2 x)$ and defined
\beq
\phi_\epsilon (\sigma) \equiv  {}_2 F_1 \left( \frac{1}{2}, 1 - \frac{\epsilon}{2}, \frac{3}{2}, - \sigma^2 \right).
\eeq

We are interested in the behavior of $\Pi(x)$ as $x \rightarrow 0$ at fixed, finite $\epsilon$.
For this, we need the large $|\sigma|$ expansion
\beq
\phi_\epsilon (\sigma) = 
\mathcal{B}_\epsilon |\sigma|^{-1} - \frac{1}{(1-\epsilon)} |\sigma|^{-2+\epsilon}
+ \mathcal{O}(|\sigma|^{-4 + \epsilon}),
\eeq
where
\beq
\mathcal{B}_\epsilon \equiv \frac{\sqrt{\pi} \Gamma(1/2-\epsilon/2)}{2 \Gamma(1-\epsilon/2)}.
\eeq
Now we can write for $\Pi(x)$ as $x \rightarrow 0$
\bea
\Pi (x) 
&\approx&  x \int_{-1/(2x)}^{\infty} d \sigma \Biggl[ \sigma \phi_\epsilon (\sigma) - 
\mathcal{B}_\epsilon + \frac{(1/x+\sigma)^{-1+\epsilon}}{(1-\epsilon)} \Biggr]^4 \nn
&\approx&  x \int_{-1/(2x)}^{\infty} d \sigma \left[ \sigma \phi_\epsilon (\sigma) - 
\mathcal{B}_\epsilon \right]^4 + \frac{4 x}{(1-\epsilon)} \int_{-1/(2x)}^{\infty} d \sigma \left[ \sigma \phi_\epsilon (\sigma) - 
\mathcal{B}_\epsilon \right]^3 \, (1/x+\sigma)^{-1+\epsilon}\nn
&\approx&  x \int_{-1/(2x)}^{0} d \sigma \left[ \sigma \phi_\epsilon (\sigma) - 
\mathcal{B}_\epsilon \right]^4 + x \int_0^{\infty} d \sigma \left[ \sigma \phi_\epsilon (\sigma) - 
\mathcal{B}_\epsilon \right]^4 - \frac{32 \mathcal{B}_\epsilon^3 x}{(1-\epsilon)} \int_{-1/(2x)}^{0} d \sigma  \, (1/x+\sigma)^{-1+\epsilon}\nn
&\approx & 8\mathcal{B}_\epsilon^4 
- \frac{32 \mathcal{B}_\epsilon^3 2^{-\epsilon}}{\epsilon(1-\epsilon)} x^{1-\epsilon}
+ x \int_{-1/(2x)}^{0} d \sigma \left( \left[ \sigma \phi_\epsilon (\sigma) - 
\mathcal{B}_\epsilon \right]^4 - 16 \mathcal{B}_\epsilon^4 + \frac{32 \mathcal{B}_\epsilon^3}{(1-\epsilon)} (-\sigma)^{-1+ \epsilon} \right) \nn
&~&~~~~~
+ x \int_0^{\infty} d \sigma \left[ \sigma \phi_\epsilon (\sigma)  - 
\mathcal{B}_\epsilon \right]^4 - \frac{32 \mathcal{B}_\epsilon^3 (1 - 2^{-\epsilon})}{(1-\epsilon)\epsilon} x^{1-\epsilon} \nn
&\approx & 8\mathcal{B}_\epsilon^4 
- \frac{32 \mathcal{B}_\epsilon^3 }{\epsilon(1-\epsilon)} x^{1-\epsilon}
+ \mathcal{D}_\epsilon \, x \label{smallG} 
\eea
where
\bea
\mathcal{D}_\epsilon &=& \int_{-\infty}^{0} d \sigma \left( \left[ \sigma \phi_\epsilon (\sigma) - 
\mathcal{B}_\epsilon \right]^4 - 16 \mathcal{B}_\epsilon^4 + \frac{32 \mathcal{B}_\epsilon^3}{(1-\epsilon)} (-\sigma)^{-1+ \epsilon} \right) + \int_0^{\infty} d \sigma \left[ \sigma \phi_\epsilon (\sigma)  - 
\mathcal{B}_\epsilon \right]^4 \nn
&=& \frac{4 \pi^3}{\epsilon} +  \left( - 6 \gamma_E + 4 \ln (2) - 6 \psi (1/2) \right)\pi^3 + 6 \pi \zeta (3) + \mathcal{O}(\epsilon),
\eea
where $\psi$ is the digamma function.
We have verified numerically that the small $x$ expansion for $G(x)$ in Eq.~(\ref{smallG}) holds accurately for small values of $\epsilon$.

We can now insert the expansion (\ref{smallG}) in Eq.~(\ref{IG}) and obtain the singular terms
in $\mathcal{I}(\epsilon)$ as $\epsilon \rightarrow 0$:
\bea
\mathcal{I} (\epsilon) &=& \frac{\mathcal{A}_\epsilon}{24} \left( - \frac{32 \mathcal{B}_\epsilon^3}{2 \epsilon^2 (1 - \epsilon)} + \frac{\mathcal{D}_\epsilon}{3 \epsilon} \right) \nn
&=& - \frac{\pi^2}{9 \epsilon^2} + \frac{\mathcal{E}}{\epsilon}
\eea
where
\bea
\mathcal{E} &=& \frac{6 \zeta (3)+\pi ^2 \left(-5+\ln (64)+3 \psi
   \left({1}/{2}\right)-\psi
   \left({3}/{2}\right)\right)}{18 
   } \nn
&=& - 3.310360722 \ldots   
\eea
Now we use the lower order result for $G(\tau)$ in Eq.~(\ref{tau1}), insert the above result for $\mathcal{I} (\epsilon)$ in Eq.~(\ref{3loop}), and evaluate
at $\tau \mu=1$, and keep only poles in $\epsilon$,  to obtain
\bea
\frac{G (\tau)}{G_0 (\tau)} &=& \exp \left( - \frac{\gamma^2 Z_\gamma^2}{\epsilon (1-\epsilon)}  \right) \left[ 1 - g \gamma^4 \mathcal{I}(\epsilon) + \mathcal{O}(g^2)\right] \nn
&=& \exp \left( - \frac{\gamma^2}{\epsilon} \right) \left[ 1 - g \gamma^4 \left( \frac{2 \pi^2}{9 \epsilon^2} + \frac{1}{\epsilon} \left(\mathcal{E} +
\frac{\pi^2}{3} \right)  \right) + \mathcal{O}(g^2)\right] \nn
&=& \exp \left( - \frac{\gamma^2}{\epsilon} \right) \left[ 1 - g \gamma^4 \left( \frac{2 \pi^2}{9 \epsilon^2} - \frac{0.020492588211}{\epsilon}  \right) + \mathcal{O}(g^2)\right]. \label{Cres}
\eea
Demanding cancellation in poles for the renormalized fermion $\psi_R$, we obtain Eq.~(\ref{Zhres}).

\section{Self energy renormalization}
\label{app:selfen}

This appendix will carry out a computation equivalent to that in Appendix~\ref{app:3loop}, but using a Dyson formulation of the 
fermion propagator in frequency space.
In this formulation, we introduce  the self energy, $\Sigma$, defined by
\beq
G(\omega) = \frac{1}{-i \omega + \lambda - \Sigma (\omega)}.
\eeq
Then, at order $\gamma^2$, the self energy is 
\beq
\Sigma_\gamma (\tau) = \gamma_0^2 \, \theta (\tau) \, D_0 (\tau) e^{- \lambda \tau}
\eeq
So we have 
\bea
\Sigma_\gamma 
( \omega)  &=& \mu^{\epsilon} \gamma^2 Z_\gamma^2 \int_0^{\infty} \frac{d \tau}{\tau^{2-\epsilon}} e^{- (\lambda - i \omega)\tau} \nn
&=&  \mu^{\epsilon} (\lambda - i \omega)^{1-\epsilon} \gamma^2 Z_\gamma^2 \,  \Gamma(-1 + \epsilon) \nn
&=&  \mu^{\epsilon} (\lambda - i \omega)^{1-\epsilon} \gamma^2 Z_\gamma^2 \left( - \frac{1}{\epsilon} - 1 + \gamma_E + \ldots \right) \label{S1}
\eea 
So in minimal subtraction, we have at order $\gamma^2$, 
\beq
Z_h = 1 - \frac{\gamma^2}{\epsilon}
\label{Z1a}
\eeq
which agrees with Eq.~(\ref{Z1}).

\begin{figure}
\includegraphics[width=2in]{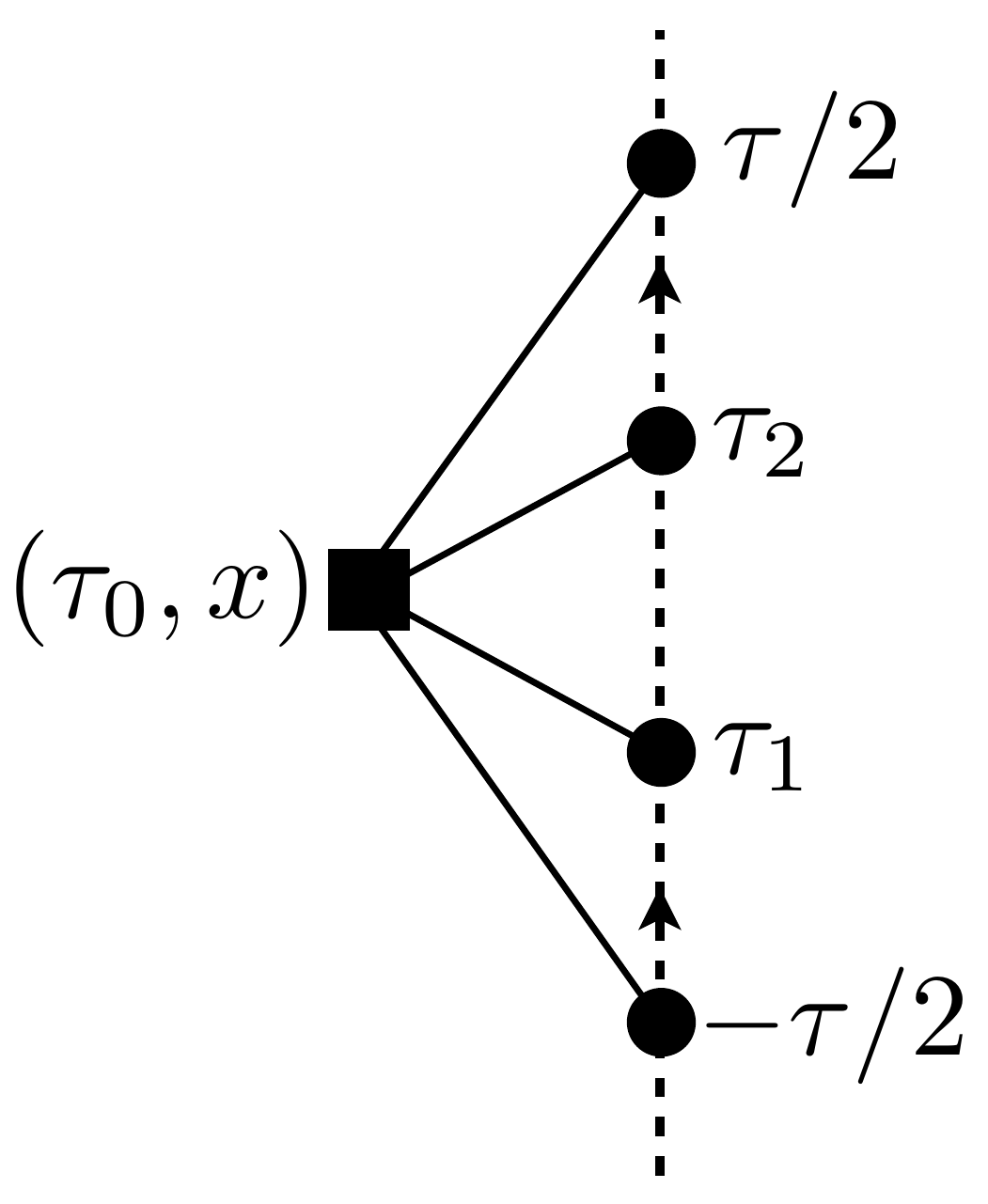} 
\caption{Feynman diagrams for the fermion self-energy at order $g$.
}
\label{fig:feyn3}
\end{figure}
We now turn to the terms of order $g$, where we need to compute the 3-loop self-energy term. This is given by the Feynman diagram
in Fig.~\ref{fig:feyn3} and leads to an expression very similar to that in Eq.~(\ref{V1})
\bea
\Sigma_g (\tau)  &=& - \gamma_0^4 g_0 \widetilde{S}_{d+1}^4  \int d^d x \int_{-\infty}^{\infty} d \tau_0 
\frac{1}{\{[x^2 + (\tau_0 + \tau/2)^2][x^2 + (\tau_0 - \tau/2)^2]\}^{(d-1)/2}} \nn
&~&~~~~~~~~~~~~~~~~\times \left[ \int_{-\tau/2}^{\tau/2}  d \tau_1 \int_{-\tau_1}^{\tau/2}  d \tau_2 \,
\frac{1}{\{[x^2 + (\tau_0 - \tau_1)^2][x^2 + (\tau_0 - \tau_2)^2]\}^{(d-1)/2}} \right] \nn
 &=& - \frac{\gamma_0^4 g_0 \widetilde{S}_{d+1}^4}{2} \int d^d x \int_{-\infty}^{\infty} d \tau_0 
\frac{1}{\{[x^2 + (\tau_0 + \tau/2)^2][x^2 + (\tau_0 - \tau/2)^2]\}^{(d-1)/2}} \nn
&~&~~~~~~~~~~~~~~~~~~~~~~~~~~~~~~~~~~\times \left[ \int_{-\tau/2}^{\tau/2}  d \tau_1 \,
\frac{1}{[x^2 + (\tau_0 - \tau_1)^2]^{(d-1)/2}} \right]^2
\eea
The second expression differs from Eq.~(\ref{V1}) primarily by the square over the integral over $\tau_1$. Rescaling to pull out the $\tau$ dependence, we now have 
\bea
\Sigma_g (\tau)  &=& - \frac{\mu^{3 \epsilon}}{\tau^{2-3\epsilon}} \, \frac{\gamma^4 g \widetilde{S}_{d+1}^2}{2 S_{d+1}} \int d^d x \int_{-\infty}^{\infty} d \tau_0 
\frac{1}{\{[x^2 + (\tau_0 + 1/2)^2][x^2 + (\tau_0 - 1/2)^2]\}^{(d-1)/2}} \nn
&~&~~~~~~~~~~~~~~~~~~~~~~~~~~~~~~~~~~\times \left[ \int_{-1/2}^{1/2}  d \tau_1 
\frac{1}{[x^2 + (\tau_0 - \tau_1)^2]^{(d-1)/2}} \right]^2
\eea
Now the Fourier transform to $\Sigma_g (\omega)$ will yield a pole in $\epsilon$ from the $1/\tau^{2-3\epsilon}$ term, just as in Eq.~(\ref{S1}). 
However the integrals of $x,\tau_0, \tau_1$ yields an additional pole in $\epsilon$, and so we cannot set $\epsilon=0$ in the integrand yet.
Evaluating the Fourier transform and the integral over $\tau_1$, we obtain
\beq
\Sigma_g (\omega) =  - \mu^{3 \epsilon} (\lambda - i \omega)^{1-3 \epsilon} \, \gamma^4 g \mathcal{A}_\epsilon \int_0^{\infty}  x^\epsilon dx \int_0^{\infty} d \tau_0 \frac{[ \Phi(x, \tau_0) ]^2}{\{[x^2 + (\tau_0 + 1/2)^2][x^2 + (\tau_0 - 1/2)^2]\}^{1-\epsilon/2}},
\eeq
where
\bea
A_\epsilon &\equiv& \mathcal{A}_\epsilon \, 2 \Gamma(-1+3 \epsilon) \nn
&=&  - \frac{2}{3 \pi\epsilon} + \ldots,
\eea
has a simple pole at $\epsilon=0$.
Now we  write
\beq
\Sigma_g (\omega) =  - \mu^{3 \epsilon} (\lambda - i \omega)^{1-3 \epsilon} \, \gamma^4 g \, A_\epsilon \, B_\epsilon
\label{S2}
\eeq
where
\beq
B_\epsilon =  \int_0^{\infty}  x^{-3+3\epsilon} dx \, 
H_\epsilon (x)
\label{S3}
\eeq
and
\bea
H_\epsilon (x) &=&  \int_{-1/(2x)}^{\infty} d \sigma \Biggl[ \sigma \phi_\epsilon (\sigma) - \left( \frac{1}{x} + \sigma \right) \phi_\epsilon \left( \frac{1}{x} + \sigma \right) \Biggr]^2 \nn
&~&~~~~~~~~~~~\times \frac{1}{\{[1 + (1/x+\sigma)^2][1 + \sigma^2]\}^{1-\epsilon/2}}. \label{H1}
\eea

We now need the expansion of $H_\epsilon (x)$ at small $x$. 
\beq
H_\epsilon (x \rightarrow 0) = x^{2-\epsilon} \int_{-\infty}^{\infty}  d \sigma \left[ \sigma \phi_\epsilon (\sigma) - \mathcal{B}_\epsilon \right]^2 \frac{1}{[1 + \sigma^2]^{1-\epsilon/2}} \equiv x^{2-\epsilon} D_\epsilon 
\eeq
where 
\beq
D_0 = \frac{\pi^3}{3}.
\eeq
Then we can construct the behavior of $B_\epsilon$ at small $\epsilon$ from Eqs.~(\ref{S3},\ref{H1})
\beq
B_\epsilon = \frac{\pi^3}{6 \epsilon} + 1.694 + \mathcal{O}(\epsilon)
\eeq
The $\mathcal{O}(1)$ term above was obtained by numerical evaluation of the integrals.

So we see from Eqs.~(\ref{S2},\ref{S3}) that the 
self energy evaluates to
\beq
\Sigma_g (\omega) =  - \mu^{3 \epsilon} (\lambda - i \omega)^{1-3 \epsilon} \, \gamma^4 g 
\left[ - \frac{\pi^2}{9 \epsilon^2} - \frac{1.41144}{\epsilon} + \mathcal{O}(1) \right]
\label{S4}
\eeq
Combining this with the lower order result $\Sigma_\gamma$ in Eq.~(\ref{S1})
at $\mu (\lambda - i \omega) = 1$, while keeping only poles in $\epsilon$, we obtain
\bea
\Sigma (\omega)  &=&  (\lambda - i \omega) \left[ \gamma^2 Z_\gamma^2 \Gamma(-1+\epsilon) + \gamma^4 g \left( \frac{\pi^2}{9 \epsilon^2} + \frac{1.411}{\epsilon}\right) + \mathcal{O}(g^2) \right] \nn
&=&  (\lambda - i \omega) \left[ - \frac{\gamma^2}{\epsilon} - \gamma^4 g \left( \frac{2 \pi^2}{9 \epsilon^2} -  \frac{0.0205}{\epsilon} \right) + \mathcal{O}(g^2) \right].
\eea
Note the excellent agreement of the renormalization factor with the exact results in Eq.~(\ref{Cres}).

\end{document}